# Trustworthy content push

Nicolai Kuntze, Andreas U. Schmidt
Fraunhofer-Institute for Secure Information Technology SIT
Rheinstrasse 75
64295 Darmstadt, Germany
Email: {nicolai.kuntze,andreas.u.schmidt}@sit.fraunhofer.de

*Abstract*— Delivery of content to mobile devices gains increasing importance in industrial environments to support employees in the field. An important application are e-mail push services like the fashionable Blackberry. These systems are facing security challenges regarding data transport to, and storage of the data on the end user equipment. The emerging Trusted Computing technology offers new answers to these open questions.

## I. INTRODUCTION

The share of workers occupied with 'nomadic' tasks in mobile contexts is constantly growing over the last years. These workers depend on infrastructures for easy and swift access to essential data. This has contributed to the success of data push services like RIM's Blackberry, which conquers the market quickly. E-Mail push is one basic value proposition aiming at high availability and ease of use. Push services are characterised by the ability to notify end users of new content. For an e-mail service the end user device is activated by the central mail server, receives the new mail, and alerts the user. Some providers have extended their service range to enable access to company databases and implement loosely coupled work-flows incorporating nomadic workers.

Due to the high potential value of the exchanged data these systems are threatened by, maybe even professional attackers, raising the requirement to protect the distribution of pushed data. The technical problems entailed by this requirement are in the focus of the present contribution. Pertinent security concerns can be grouped in two main areas. First, data has to be protected on the way to the device. During the transport the main concern is to maintain confidentiality *viz* prevent eavesdropping. Second, after the data is delivered to the device the data has to be protected against unauthorised access. The latter problem is of practical importance in use cases like the mentioned e-mail push, but also for ordinary e-mail, SMS delivery, and other data synchronisation processes between a central data base and a mobile device. Current approaches to this challenge are predominantly using software tokens, e.g., PKCS#7 as containers of credentials used to secure message transport and storage. Software solutions however suffer from the drawback that an attacker can extract the keys from memory during the encryption or decryption of a data block. Smart cards are one approach to this problem but in the mobile domain no standard has managed to become prevalent.

The security problems of push services are adjacent to general protection of content in transmission. Protection of mobile content is a major use case for Trusted Computing [1].

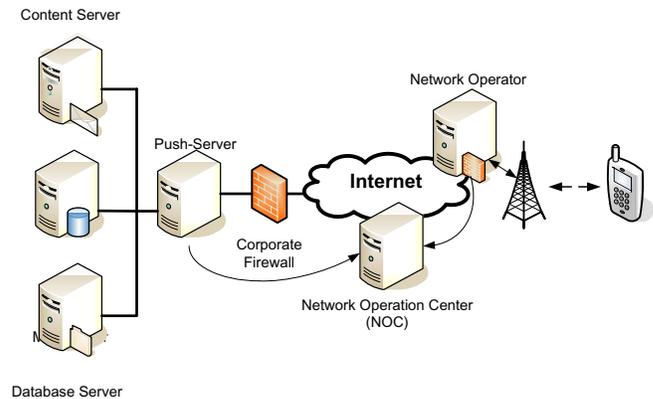

Fig. 1.   A centralised architecture

It therefore stands to reason to use Trusted Computing (TC) for a novel approach to push service security. We present two technical scenarios relying on Trusted Computing to mitigate the sketched security problems. The following section introduces basics of content push architectures and discusses their security issues. Section III lays out the basic facts about TC needed in the following, and Section IV exhibits our concepts proper. Section V contains conclusions.

## II. CONTENT PUSH ARCHITECTURES

The idea behind a content push architecture is to inform a user if new data is available without necessitating any user interaction. The basic method for this has been formulated for instance in the standards of the Open Mobile Alliance (OMA, see [2]).

Notifying a user can be done using either a centralised or decentralised approach. Figure 1 illustrates the former. In the centre of this architecture a Network Operation Center (NOC) performs all activities regarding the communication to the mobile devices. The data destined for the mobile devices are stored in sources like mail servers. These sources excite the push server and deliver the data. Due to this activation the push server either requests a communication line to the mobile device managed by the NOC or delivers the data to the NOC which in turn stores the data until they are handed to the mobile device. From a company's view the management costs are low in this scenario, as there is no additional effort needed to maintain, e.g., a special firewall configuration.

Figure 2 shows the decentralised counterpart where no central server is in charge to communicate to the attached devices. This scheme differs from the one above in that the communication between push server and mobile device is direct. For a company using this scheme it is necessary to enable device access over, e.g., the Internet to servers behind a company firewall. This requires a specially hardened network topology like a Demilitarized Zone (DMZ) housing the push server or a proxy server to mediate access to the push infrastructure inside the corporate network.

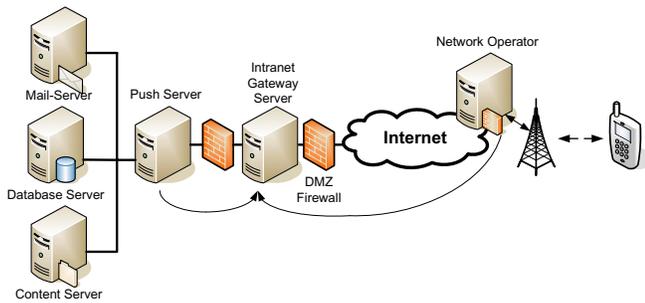

Fig. 2. A decentralised architecture

The centralised scheme is the prevalent architecture used by the market leader. Nevertheless, it raises some security concerns in particular with regard to availability and privacy. Denial of service attacks aiming at the service level of a NOC lead in the worst case to a complete service breakdown. High value and visibility of the target make corresponding attacks very likely. If we take a malicious service provider into consideration privacy concerns are added to the general ones regarding transport security. Potentially the NOC can access every message which is sent to a mobile device, allowing for information leakage. First and foremost the message content could be extracted and disclosed. Moreover, analysis of the collaboration between active users becomes possible. Both attacks have to be treated in the protocol design to enable end-to-end security and to conceal all sensitive information. It should be noted that end-to-end security is to be established between the push server and the mobile device in contrast to well known protocols like PGP or S/MIME.

Beside the requirements on the infrastructure the mobile device itself must also be examined from a security perspective. All data transferred to the end user device is stored there. An attacker can gain access to this data by theft or penetration attacks (e.g. by Bluetooth), or social engineering. Most of these security considerations are not considered in existing standards. For instance the document [2, Section 11] regards mainly authentication issues.

### III. TRUSTED COMPUTING ESSENTIALS

Trusted computing uses a hardware anchor as a root of trust and is now entering the mobile domain with the aim to provide a standardised security infrastructure. Trust in the context of TC means (as defined by the TCG) that an entity always behaves in the expected manner for the intended purpose. The trust anchor, called Trusted Platform Module (TPM), offers various functions from two main areas. First the ability of the TPM to create, store, and use asymmetric key pairs and secondly, the ability to provide an assertion to the exterior about the system state and the integrity of the system. Each TPM is bound to a certain environment and together they form a trusted platform (TP) from which the TPM cannot be removed. Through the TPM the TP has access to a cryptographic engine and a protected storage. Each physical instantiation of a TPM has a unique identity embodied in an Endorsement Key (EK) which is created at manufacture time. This key is used as a base for secure transactions as the Endorsement Key Credential (EKC) asserts that the holder of the private portion of the EK is a TPM conforming to the TCG specification. The EKC is issued as well at production time and the private part of the key pair never leaves the TPM. There are other credentials stating the conformance of the TPM and the platform, which are not of importance here. Before a TPM can be used a take ownership procedure must be performed in which the usage of the TPM is bound to a certain user. The following technical details are taken from [3].

For the TPM to issue an assertion about the system state, two protocols are available for the process called attestation. As the uniqueness of every TPM leads to privacy concerns, they provide pseudonymity, respectively, anonymity. Both existing attestation protocols rest on Attestation Identity Keys (AIKs) which are placeholders for the EK. An AIK is a 1024 bit RSA key the private portion of which is sealed inside the TPM. The simpler protocol of Remote Attestation (RA) offers pseudonymity by introducing a trusted third party, the privacy CA (PCA, see [4]), which issues a credential stating that the respective AIK is generated by a sound TPM within a valid platform. This credential together with the AIK can therefore be used as an identity for this platform. The system state is measured by a reporting process with the TPM as its central reporting authority receiving the measurement values and calculating a unique representation of the state using hash values. For this, the TPM has several Platform Configuration registers (PCR). Beginning with the system boot each component reports a measurement value, e.g., a hash value over the BIOS, to the TPM and stores it in a log file. During RA, the communication partner which wants to establish trust in the TP and acts as verifier receives this log file and the corresponding PCR value. The verifier can then decide if the device is in a configuration which is trustworthy from his perspective. Apart from RA, the TCG has defined Direct Anonymous Attestation. This involved protocol is based on a zero knowledge proof but due to certain constraints of the hardware it is not implemented in current TPMs.

Beside the attestation methods TC offers a concept to bind blobs of data to a single instantiation and state of a TPM, which is essential for the concept presented below. The operation takes the data blob that is the result of a command and decrypts it for export to the user. The caller must authorise the use of the key that will decrypt the incoming blob. In consequence this data blob is only accessible if the platform is in the namely state which is associated with the respective PCR value.

A mobile version of the TPM is currently being defined by the TCG's Mobile Phone Working Group [5]. This so-called Mobile Trusted Module (MTM) differs significantly from the TPM of the PC world and is in fact more powerful in some respects. In particular, it contains a built-in verifier for attestation requests, substituting partly for an external PCA. Both TPM and MTM are a solid basis for the push service security concept we now present.

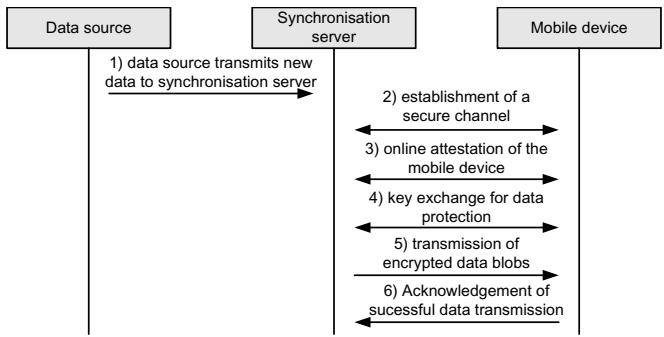

Fig. 3. Scenario 1 based on blob sealing

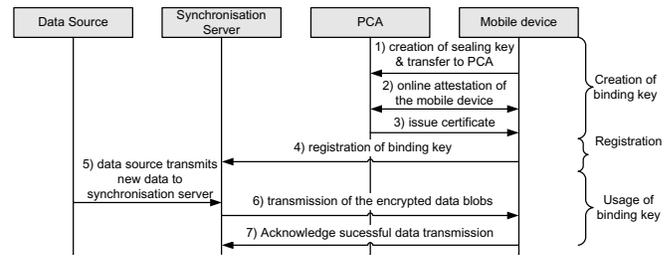

Fig. 4. Scenario 2 based on key sealing

## IV. CONCEPTS FOR TRUSTED CONTENT PUSH

Based on blob binding as presented before we develop a basic scenario to protect the data which is exchanged between a data source and a mobile device. Figure 3 exemplifies the simple protocol between the actors of a potential uses case. The data source (e.g., a mail server) receives new data which are to be synchronised with the mobile device. Step 1 signals the synchronisation server which controls the synchronisation process. This can be implemented as a push (data source activated) or as a pull scheme (the synchronisation server periodically polls for new data). The synchronisation server has basically the tasks to locate the mobile device by determining its unique ID or address, to establish a communication, and to control the synchronisation process. Our first concept is only concerned with the security of the data exchange, as there are various established protocols standardising the synchronisation process, e.g., OMA data synchronisation (OMA DS). In Step 2 of Figure 3 a secure channel between the synchronisation server and the mobile device is established. This can be performed employing known methods such as Transport Layer Security (TLS). After this, an attestation of the platform is required to testify the mobile device and to proof the integrity of the platform (Step 3). Security of the following steps relies on this assertion of platform trustworthiness. Step 4 performs a key exchange so that the data can be encrypted independently of channel security. This step is optional if the secure channel is strong enough and considered reliable. After this, the data is transferred to the mobile device (Step 5). The device receives the data and stores them in sealed data blobs (standardised by TCG [3, Chapter 12]). As the result of this, the data is only accessible in the chosen state of this unique device.

This basic approach suffers from the drawback of a high latency produced by the Steps 2-4. Especially remote attestation creates a high computational load and produces some traffic.

A solution to this problem is a public/private key scheme. We now refine the base concept accordingly, presenting our second concept exhibited in Figure 4. The synchronisation server now encrypts the data with a public key. The usage of the corresponding private key should be restricted by a PCR value as it is known from the first scenario. To grant trust in this public key the synchronisation server requires a certificate issued by this particular TPM or a third party stating that this public key corresponds to a private key both created by a valid TPM on a TP. Also, the particular PCR value must be enclosed in this certificate. This concept is very close to attestation identity keys (AIKs) which are used in TC for platform attestation. The AIK privacy CA which certifies these platform keys can be used here to issue certificates augmented by the information of the PCR value. This variant of an AIK is further called a *binding key*. In contrast to an AIK its aim is not to provide pseudonymity but a way to encrypt data at the side of the sender which can only be used in a predefined state on the receiver side (this feature could also be used to establish a lightweight DRM mechanism). By requiring a certain PCR value, the activity of a certain application on the device can be enforced as it is necessary that this application was reported to the TPM. In the example of an e-mail push service the presence of an e-mail application is required in a certain configuration and a well defined environment.

The top-level protocol for creation and usage of binding keys is presented in Figure 4. The data source again offers (by either push or pull) new data. The protocol is parted in three stages. Stage 1 is concerned with the key creation and the issuing of the certificate. In Stage 2 the key is registered at the side of the synchronisation server. In the final stage the key is used for its purpose. Step 1 (from Stage 1) transmits the public portion of a key pair to the PCA. Additionally several other certificates are transmitted. In Step 2 an online attestation is performed which assures the status of the mobile device and extracts the actual PCR values. Using this value a certificate is created stating that the key originates from a trusted platform and that it is useable if and only if the platform equals a certain state. This certificate is transmitted in Step 3 to the mobile device. Step 4 describes that the binding key and the corresponding certificate are transmitted to the synchronisation server. Herein it is necessary to register this data for use by a certain user. This process depends on the particular use case and is not in the scope of this paper. Steps 1 to 4 only have to be performed once during the roll out of the device or the take ownership of it by the user. In Step 5 the data is transmitted to the synchronisation server. The synchronization server can now encrypt the data with the binding key or maybe with a hybrid encryption scheme based on this key. A previous key exchange is not required. As the key is bound to a certain state it is also not required to check if the device is in this state as neither the device nor any other entity can unveil the content of the data. This takes place in Steps 5 and 6.

Let us briefly discuss efficiency matters, keeping in mind that the analysis must remain hypothetical since we do not have a real-world implementation. Comparing the two schemes it becomes obvious that the first is much more flexible in its application than the second one. The server can decide each time if the particular device can be considered as trustworthy. Binding the transmitted data to a defined system state leads to a very restricted application scenario as the system state is not necessarily static. The mobile version of the TPM, the MTM, introduces some new concepts, which might be of use with respect to this problem. The standard employs an abstraction layer by the definition of so-called trusted engines. The described functionality (of model 2) can now be located in such an engine. This engine controls the complete handling, including de/encryption and display of pushed content and thus comprises a trustworthy system compartment for this sole purpose. The rest of the system can run freely, adding much flexibility to our core concepts.

With respect to performance, the use of TC in the described way has its advantages and drawbacks. On the one hand, the hardware capabilities of the TPM or MTM add considerable power to existing systems for instance for the physical generation of random numbers. By design the MTM/TPM operates independently and in parallel to the system's CPU. On the other hand, key cryptographic tasks used in our concept are expensive. In particular the generation of AIKs is rather slow and may take, together with remote attestation, some 15 seconds. Though this restriction is partly due to the "historical" binding of the TPM to the low pin count (LPC) bus of the PC, and may even not apply to the mobile version, TCG implementers a dedicated pre-fetch mechanism for AIK generation to accelerate a TPM's response to AIK requests. For our concept the main latency introduced by TC is by remote attestation, which may take up to 8 seconds (always at LPC bus cycles). If we assume that even a new AIK is used for every push operation, as in a ticket system which we described elsewhere [6], then the overall latency might add up to 30 seconds. Push services typically have a relatively low message frequencies for a single addressee, so that the concept seems still feasible, even with current technology. Otherwise it would still be possible to perform a bulk message transfer.

## V. CONCLUSIONS

The presented solutions provide both a feasible way to synchronise data and to transport them securely from one entity to another. As shown in the second concept, one can reduce the computational costs for the server and the device as some of the steps from the first concept can be omitted by a cunning use of the Trusted Computing concepts of privacy CAs and AIKs. The presented methods represent a strong base for a lightweight DRM system based on TC. Certain data can only be decoded and used precisely if the device is in the desired trustworthy state. It should be noted that our application uses TC in a way different from DRM, which is often considered as the sole use for TC. Both applications bind the economic value to a particular instantiation of the TPM. If this trust anchor breaks, only a limited damage can occur as the damage is restricted in space and timeto a single data synchronisation in a push service or submission of a single reputation. In contrast, if a single TPM in a DRM system breaks, the protected digital good can be converted into an unprotected version which can be freely distributed on a large scale, causing heavy monetary losses to its owner.

AIKs lend themselves to more applications than blob sealing only. Since they are generated by the trusted platform itself and are available in principle in unlimited number, they can be used for various different purposes. This can go so far as to provide a basis for a full-fledged identity management. In [6] we have shown how to implement a pseudonymous ticket system based on AIKs. The papers [7], [8] exhibit some further TC uses in the mobile domain.

It is interesting to note that introducing TC yields a secondary user bound identification token. Due to the take ownership procedure of the TPM it is bound to a certain user. Therefore it is in its function very similar to a SIM as it is also possible to migrate the relevant parts from one TPM to the next. One may ask whether two different identification tokens will survive in future TC-enabled mobile devices.